\documentclass{PoS}
\usepackage{placeins}
\usepackage{amsmath}
\usepackage[numbers,sort&compress]{natbib}
\usepackage{booktabs}
\title{Higgs and the electroweak precision observables in the MRSSM}

\ShortTitle{Higgs boson and EWPO in the MRSSM}

\author{Philip Diessner$^\dag$, \speaker{Wojciech Kotlarski}{$^{\dag \S}$}
\\
\\
$^\dag$Institut f\"ur Kern- und Teilchenphysik, TU Dresden, 01069 Dresden, Germany\\
$^\S$Faculty of Physics, University of Warsaw, Pasteura 5, 02093 Warsaw, Poland\\
 
E-mail: \email{philip.diessner@mailbox.tu-dresden.de, wojciech.kotlarski@fuw.edu.pl}}

\abstract{
We briefly review recent progress in the analysis of the Higgs sector of the Minimal R-symmetric Supersymmetric Standard Model. 
Importance of the interplay between W and Higgs boson masses in constraining the parameter space of the model is shown.
}

\FullConference{Proceedings of the Corfu Summer Institute 2014 "School and Workshops on Elementary Particle Physics and Gravity",\\
		3-21 September 2014\\
		Corfu, Greece }

\begin{document}
\section{Introduction}

In recent years non-minimal SUSY models have been gaining increasing attention.
This is, on one hand, due to the discovery of a Higgs boson whose mass is close to the maximal one allowed in the MSSM, and to no discovery of SUSY particles during Run I of the LHC on the other.
Proposed extensions vary greatly, from 'simple' ones, like NMSSM, to rich and complicated like E$_6$MSSM and beyond.
One of those models is MRSSM, which is a minimal supersymmetrized Standard Model preserving R-symmetry. 
It was proposed in \cite{Kribs:2007ac}, in the context of solving the flavor problem of the MSSM.
Since the discovery of a Higgs boson by the ATLAS and CMS experiments at the LHC it became of interest if this model can accommodate it.
First answer to this question was given in \cite{Bertuzzo:2014bwa}, followed by calculation of complete one-loop corrections to Higgs and W boson masses \cite{Diessner:2014ksa} and two-loop corrections in the effective potential approximation and gauge-less limit \cite{Diessner:2015yna}.  
In this note we briefly review results presented in \cite{Diessner:2014ksa}.
%The note is organized as follows. First we describe briefly the model and than discus predictions for the W and Higgs boson masses togethe with their interplay.
\section{The MRSSM}
\label{sec:MRSSM}

Table~\ref{tab:Rcharges} lists the particle content of the MRSSM, together with R-charge assignment for the superfields.
With this assignment, R-symmetric superpotential reads
\begin{align}
\nonumber W = & \mu_d\,\hat{R}_d \cdot \hat{H}_d\,+\mu_u\,\hat{R}_u\cdot\hat{H}_u\,+\Lambda_d\,\hat{R}_d\cdot \hat{T}\,\hat{H}_d\,+\Lambda_u\,\hat{R}_u\cdot\hat{T}\,\hat{H}_u\,\\ 
 & +\lambda_d\,\hat{S}\,\hat{R}_d\cdot\hat{H}_d\,+\lambda_u\,\hat{S}\,\hat{R}_u\cdot\hat{H}_u\,
 - Y_d \,\hat{d}\,\hat{q}\cdot\hat{H}_d\,- Y_e \,\hat{e}\,\hat{l}\cdot\hat{H}_d\, +Y_u\,\hat{u}\,\hat{q}\cdot\hat{H}_u\, ,
\label{eq:superpot}
 \end{align} 
where $\hat{H}_{u,d}$ are the MSSM-like Higgs weak iso-doublets, and $\hat{S},\, \hat{T},\, \hat{R}_{u,d}$ are  the singlet, weak iso-triplet and $\hat{R}$-Higgs weak iso-doublets, respectively. 
The usual MSSM $\mu$-term is forbidden; instead the $\mu_{u,d}$-terms involving R-Higgs fields are allowed. 
The $\Lambda,\lambda$-terms are similar to the usual Yukawa terms, where the $\hat R$-Higgs and $\hat{S}$ or $\hat{T}$ play the role of the quark/lepton doublets and singlets.

Since Majorana masses for the gauginos are forbidden by the R-symmetry, appearance of gauge single, $SU(2)_L$ triplet and $SU(3)_C$ octet superfields $\hat S$, $\hat T$ and $\hat O$ is dictated by the need to write their Dirac mass terms of the form
\begin{align}
V_D \ni M_B^D \tilde{B}\,\tilde{S} + 
M_W^D \tilde{W}^a\tilde{T}^a +
M_O^D \tilde{g}^a\tilde{O}^a
+ \mbox{h.c.}\,,
\label{eq:potdirac}
\end{align}
where $\tilde B$, $\tilde W$ and $\tilde g$ are familiar MSSM Weyl fermions.

In the scalar sector, after the electroweak symmetry breaking,
(neutral) scalar components of $\hat H_d, \hat H_u, \hat{S}$ and  $\hat T$ acquire vev, which we parametrize as
\begin{align} 
H_d^0=& \, \textstyle{\frac{1}{\sqrt{2}}} (v_d + \phi_{d}+i  \sigma_{d}) \;,& 
H_u^0=& \, \textstyle{\frac{1}{\sqrt{2}}} (v_u + \phi_{u}+i  \sigma_{u}) \;, \nonumber \\ 
T^0  =& \, \textstyle{\frac{1}{\sqrt{2}}} (v_T + \phi_T +i  \sigma_T)   \;,&
S   = & \, \textstyle{\frac{1}{\sqrt{2}}} (v_S + \phi_S +i  \sigma_S)   \;.
\label{eq:vevs}
\end{align} 
Since R-Higgs bosons carry R-charge 2 their vev would spontaneously break R-symmetry leading to a massles R-axion.
Therefore we do not consider it here.

In the next section we will discus $v_T$, which is strongly constrained by the measurement of the W boson mass.

CP-even components $\{ \phi_d, \phi_u , \phi_S, \phi_T \}$ mix giving rise to 4 physical Higgs boson. 
Due to the mixing, lightest Higgs mass is always lower than in the MSSM, requiring larger radiative corrections to reach the measured value.
In sec. 4 we will show that this is indeed achievable in the MRSSM.
\begin{table}[th]
\begin{center}
\begin{tabular}{c|l|l||l|l|l|l}
%\hline
\multicolumn{1}{c}{Field} & \multicolumn{2}{c}{Superfield} &
                              \multicolumn{2}{c}{Boson} &
                              \multicolumn{2}{c}{Fermion} \\
\hline 
 \phantom{\rule{0cm}{5mm}}Gauge Vector    &\, $\hat{g},\hat{W},\hat{B}$        \,& \, $\;\,$ 0 \,
          &\, $g,W,B$                 \,& \, $\;\,$ 0 \,
          &\, $\tilde{g},\tilde{W}\tilde{B}$             \,& \, +1 \,  \\
Matter   &\, $\hat{l}, \hat{e}$                    \,& \,\;+1 \,
          &\, $\tilde{l},\tilde{e}^*_R$                 \,& \, +1 \,
          &\, $l,e^*_R$                                 \,& $\;\;\,$\,\;0 \,    \\
          &\, $\hat{q},{\hat{d}},{\hat{u}}$       \,& \,\;+1 \,
          &\, $\tilde{q},{\tilde{d}}^*_R,{\tilde{u}}^*_R$ \,& \, +1 \,
          &\, $q,d^*_R,u^*_R$                             \,& $\;\;\,$\,\;0 \,    \\
 $H$-Higgs    &\, ${\hat{H}}_{d,u}$   \,& $\;\;\,$\, 0 \,
          &\, $H_{d,u}$               \,& $\;\;\,$\, 0 \,
          &\, ${\tilde{H}}_{d,u}$     \,& \, $-$1 \, \\ \hline
\phantom{\rule{0cm}{5mm}} R-Higgs    &\, ${\hat{R}}_{d,u}$   \,& \, +2 \,
          &\, $R_{d,u}$               \,& \, +2 \,
          &\, ${\tilde{R}}_{d,u}$     \,& \, +1 \, \\
  Adjoint Chiral  &\, $\hat{\cal O},\hat{T},\hat{S}$     \,& \, $\;\,$ 0 \,
          &\, $O,T,S$                \,& \, $\;\,$ 0 \,
          &\, $\tilde{O},\tilde{T},\tilde{S}$          \,& \, $-$1 \,  \\
%\hline
\end{tabular}
\end{center}
\caption{The R-charges of the superfields and the corresponding bosonic and
             fermionic components.
        }
\label{tab:Rcharges}
\end{table}

\FloatBarrier

\section{Precision EW observables}
\label{sec:EW}
The vev $v_T$ of the $SU(2)_L$ triplet field $T^0$ in eq.~\ref{eq:vevs} breaks the custodial symmetry already at the tree level shifting the W boson mass by
\begin{equation}
  m_W^2 = \frac{1}{4} g_2^2 v^2 + g_2^2 v_T^2,
\end{equation}
where $v^2 \equiv v_u^2 + v_d^2$. Large $|v_T|$ is therefore excluded by measurement of $m_W = 80.385 \pm 0.015$~GeV \cite{Agashe:2014kda}.
Small $|v_T|$ corresponds, through tadpole equations, to large values of triplet soft-mass parameter $m_T^2$, leading to somewhat split spectrum with heaviest Higgs boson around few TeV for $|v_T|$ of few hundred MeV.

To approach experimental accuracy of 15 MeV for $m_W$ one has to include at least one-loop corrections which can be concise written in the $\overline{\text{DR}}$ scheme as (see ref. \cite{degrassi})
\begin{equation}
m_W^2 = \frac{1}{2} m_Z^2 \hat{\rho} \left [ 1 + \sqrt{1
- \frac{4 \pi \hat{\alpha}}{\sqrt{2} G_\mu m_Z^2 \hat{\rho}
(1-\Delta \hat{r}_W)}}\;\right ] 
\label{w-mass-master-formula2},
\end{equation}
where $\hat \rho$ contains only oblique while $\Delta \hat{r}_W$ both oblique and non-oblique corrections (see ref. \cite{degrassi} and \cite{Diessner:2014ksa} for the thorough discussion of this formula).
Equation \ref{w-mass-master-formula2}, although very useful for numerical evaluation of the contributions as it properly resums leading two-loop SM corrections \cite{degrassi}, does not give direct insight into importance of different contributions due to implicit cancelations between $\hat \alpha, \hat \rho$ and $\hat r_W$.

\begin{figure}
  \centering
  \includegraphics[width=0.45\textwidth]{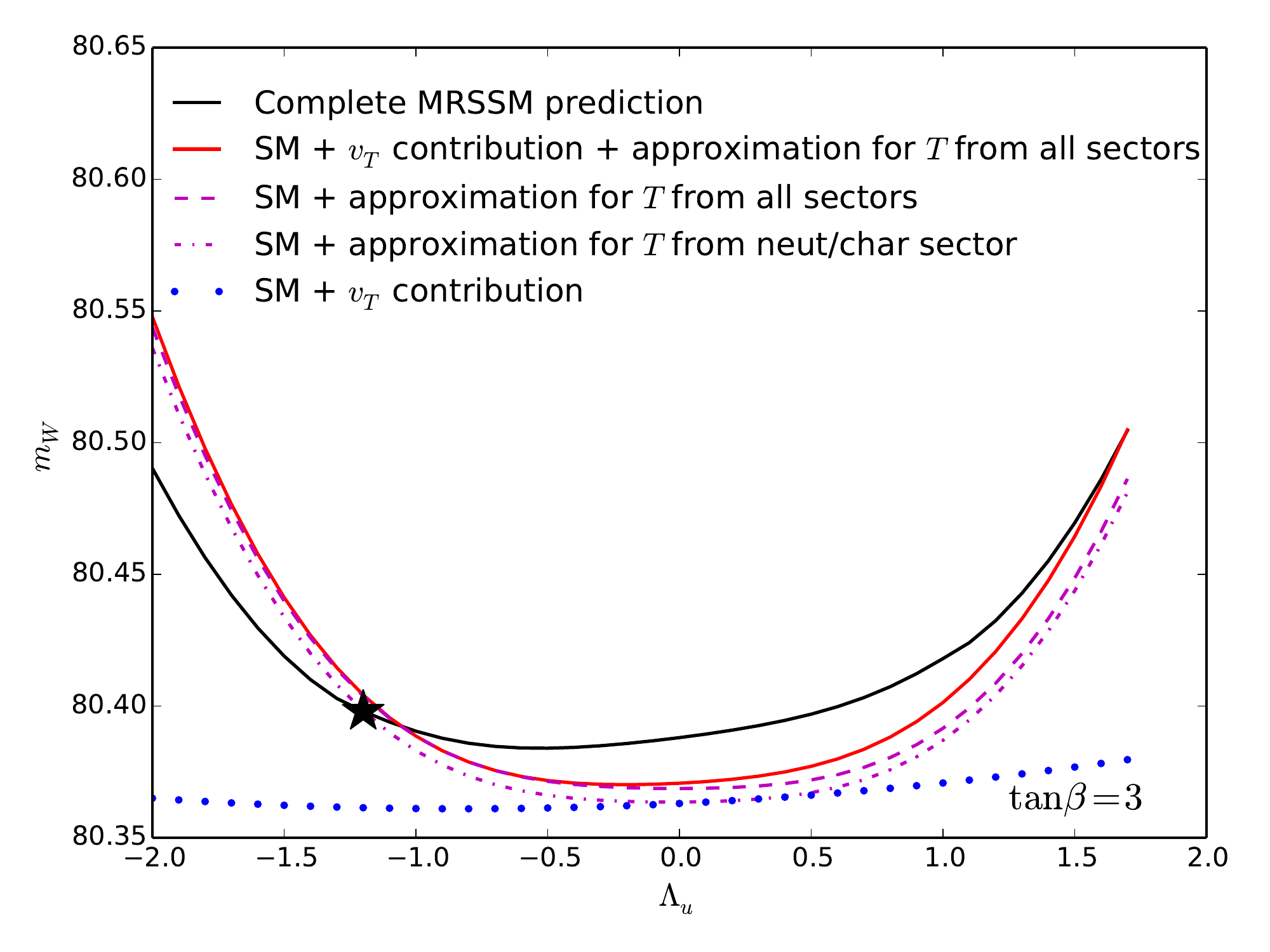}
  \caption{Comparison of the mass of the W boson depending on $\Lambda_u$, calculated using full MRSSM contributions and different approximations
for the $T$-parameter. Black stars marks the benchmark points of ref \cite{Diessner:2014ksa}.}
  \label{fig:mW_decomposition}
\end{figure}

Expanding eq.~\ref{w-mass-master-formula2} and recasting it in terms of familiar S,T and U parameters \cite{Peskin:1990zt,Marciano:1990dp,Peskin:1991sw,Kennedy:1990ib,Kennedy:1991sn,Altarelli:1990zd} we get
\begin{equation}
m_W= m_W^{\text{ref}} +\frac{\hat{\alpha} m_Z \hat{c}_W}{2(\hat{c}_W^2-\hat{s}_W^2)}\left (-\frac{S}{2}+\hat{c}^2_WT+\frac{\hat{c}^2_W-\hat{s}^2_W}{4 \hat{s}_W^2}U \right)\;.
\label{eq:mW-STU}
\end{equation}
where $m_W^{\text{ref}}$ is W boson mass as calculated in the SM. Figure \ref{fig:mW_decomposition} shows result of this decomposition for one of the benchmark points of ref. \cite{Diessner:2014ksa}. We see that the full result (given by the black line) is well approximated by the sum of tree and one-loop contributions to the $T$-parameter.
Importance of formula \ref{eq:mW-STU} comes from the fact that in many cases one can find relatively simple, concise expressions for the $T$-parameter.
Also, $T$-parameter constraints not only $m_W$, but also other EW precision observables.
For the benchmark points devised in \cite{Diessner:2014ksa}, the total contributions to the $T$-parameter were always smaller than 0.1.

\FloatBarrier
\section{Higgs mass at one loop}
\label{sec:Higgs}
\begin{figure}
  \centering
  \includegraphics[width=0.4\textwidth]{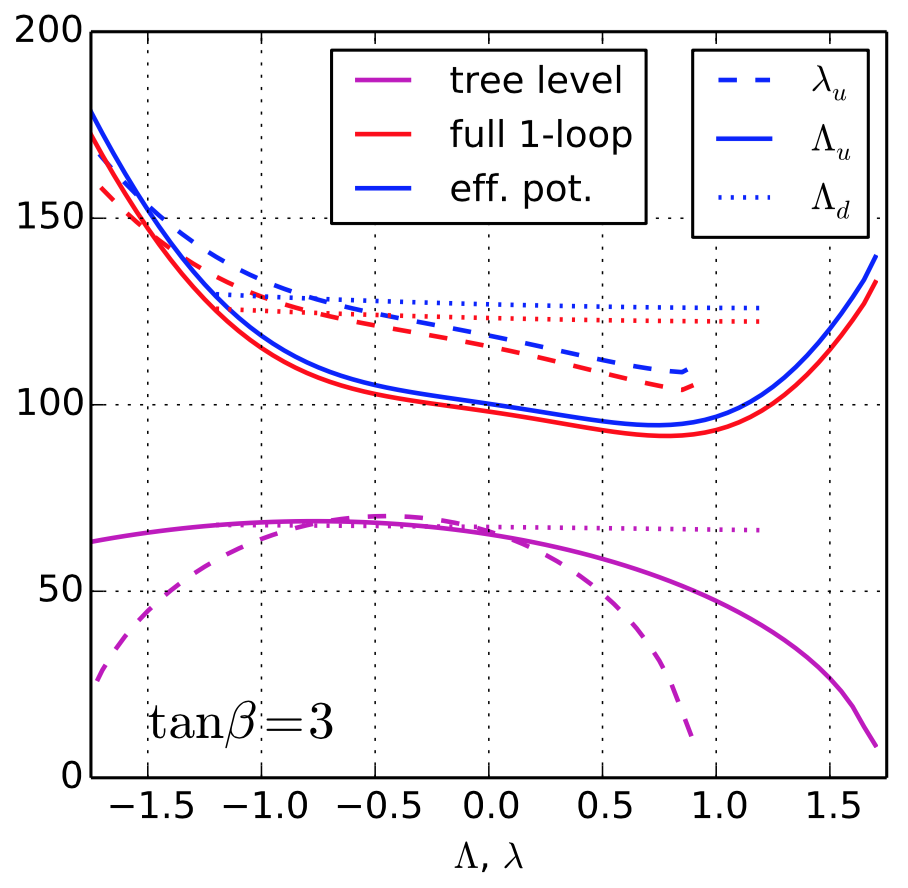}
  \caption{Comparison of the lightest Higgs boson mass calculated using the effective potential
approach (blue)  and the full one-loop calculation (red), as well as the tree-level mass (magenta).  Results are shown as functions of one of the couplings: $\Lambda_u$ (solid), $\Lambda_d$ (dots), $\lambda_u$ (dashes), 
for benchmark point 1 of ref. \cite{Diessner:2014ksa}.}
  \label{fig:eff_pot_vs_full}
\end{figure}

\begin{figure}
  \centering
  \includegraphics[width=0.5\textwidth]{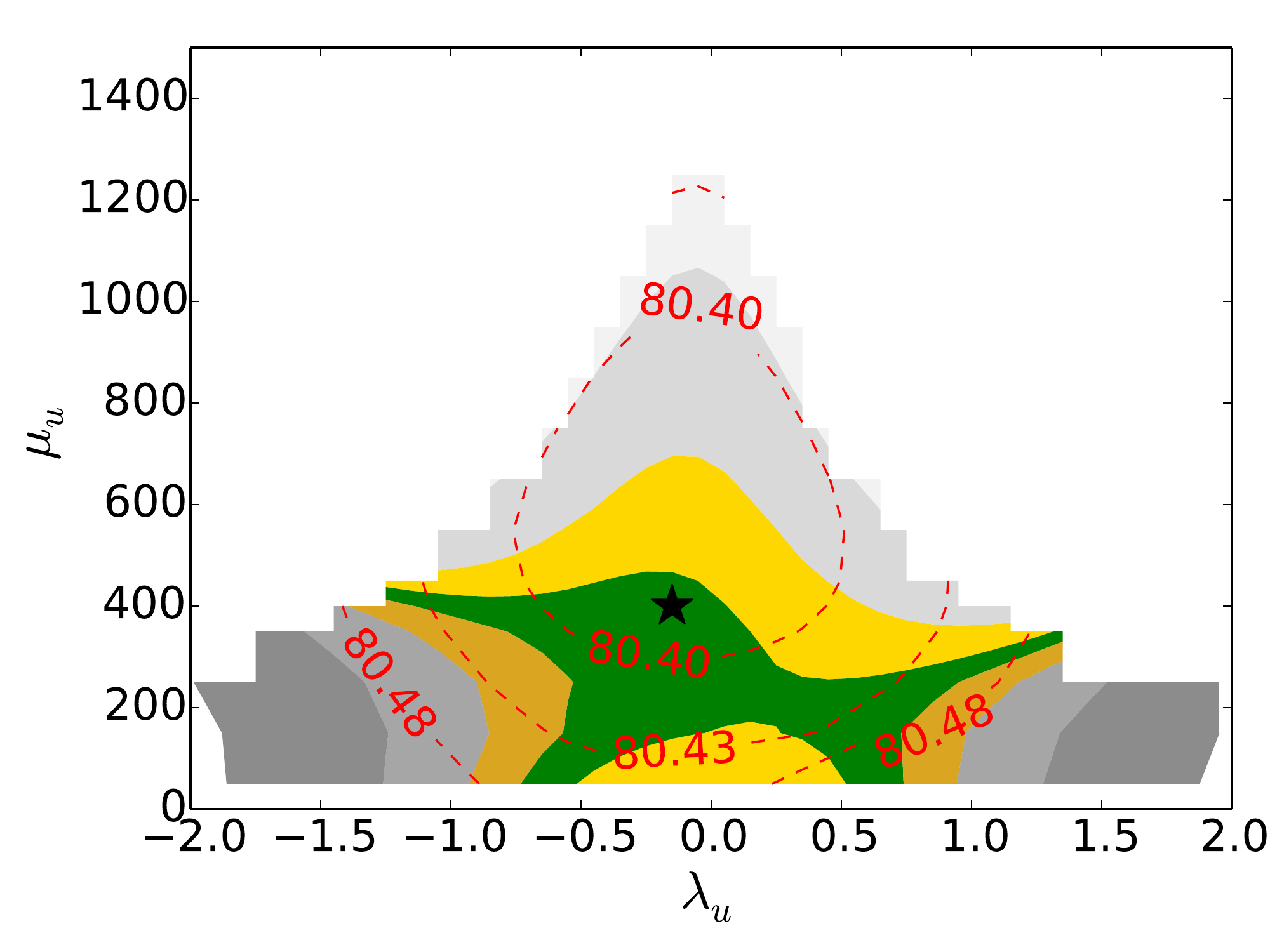}
  \caption{Interplay between Higgs and W boson masses for superpotential parameters $\lambda_u$ and $\mu_u$. Green region corresponds to $m_h = 126 \pm 2$ GeV, yellow to $\pm8$ GeV. Contours show value of $m_W$.}
  \label{fig:plot}
\end{figure}

As already pointed out, the lightest Higgs boson mass in the MRSSM suffers at the tree-level from reduction (compared to MSSM)  due to mixing with singlet and triplet states.
This can be seen from an approximate formula, when using the MSSM mixing angle $\alpha$ to diagonalize 
$\{\phi_d, \phi_u \}$ submatrix for large $m_A^2$ when $\alpha=\beta-\pi/2$, 
further assuming $\lambda=\lambda_u=-\lambda_d$, $\Lambda=\Lambda_u=\Lambda_d$, $\mu_u=\mu_d=\mu$ and $v_S\approx v_T\approx0$, which reads 
\begin{equation}
m_{H_1,\text{approx}}^2 = m_Z^2 \cos^2 2\beta - v^2 \left(
\frac{\left(g_1 M^D_B+\sqrt{2}\lambda\mu\right)^2}{4(M^D_B)^2+ m_S^2}
+
\frac{\left(g_2 M^D_W+\Lambda\mu\right)^2}{4(M_W^D)^2+ m_T^2}
\right) \cos^2 2\beta\;.
\label{eq:approx_treehiggs}
\end{equation}
Beyond the tree-level the Higgs mass receives large corrections which push its value towards the measured one. 
This is achieved with stops of mass around 1 TeV and without L-R mixing in the squark sector which is forbidden by the R-symmetry.
Figure~\ref{fig:eff_pot_vs_full} compares impact of full one-loop corrections with ones calculated in the effective potential approximation for 3 superpotential parameters: $\Lambda_u, \Lambda_d$ and $\lambda_u$.
Large values of couplings $\Lambda, \lambda$, needed to lift the mass of the lightest Higgs, could in principle cause conflict with W boson mass measurement.
This is exemplified in fig.~\ref{fig:plot} where the mentioned interplay between $m_h$ and $m_W$ predictions for one selected pair of superpotential parameters is shown.
We emphasize that the analysis of $T$-parameter contributions discussed in the previous section was done only in order to understand (approximate) functional dependence of $m_W$ on the parameters of the model.
All numerical calculations were done using eq. \ref{w-mass-master-formula2}.
As already pointed out it the abstract, it is clear from fig.~\ref{fig:plot} that W and Higgs boson masses give non-trivial constrains on the parameter space of the model. 
Nevertheless it is easy to identify regions in the parameter space which
accommodate both measurements.
\section{Conclusions}

In this work we have reviewed recent progress in the analysis of the Minimal R-symmetric Supersymmetric Standard Model. 
We showed that the model can accommodate measured Higgs boson mass while being in agreement with measured mass of the W boson.
The $\sim$125 GeV Higgs mass is obtained without stop mixing, which is forbidden by the R-symmetry, and with stops of masses below 1 TeV, which is an attractive feature, both theoretically and experimentally.
Despite that, there are still some open questions like issue of dark matter or confronting MRSSM with LHC limits from direct searches of SUSY particles.
We hope that these question will be addressed, with positive outcome, in the near feature.

\vspace{0.5cm}
\textit{In ref. \cite{Diessner:2015yna} we substantially refined the calculation by including two-loop corrections to Higgs mass in effective potential approximation and gauge-less limit.
Although important, those contributions don't change conclusions of \cite{Diessner:2014ksa}.}

\acknowledgments
Work supported in part by the 
German DFG Research Training Group 1504 and the DFG grant STO 876/4-1,
the Polish National Science Centre grants under OPUS-2012/05/B/ST2/03306
and the European Commission through the contract PITN-GA-2012-316704 (HIGGSTOOLS).


\begin{thebibliography}{99}
%\cite{Kribs:2007ac}
\bibitem{Kribs:2007ac}
  G.~D.~Kribs, E.~Poppitz and N.~Weiner,
  %``Flavor in supersymmetry with an extended R-symmetry,''
  Phys.\ Rev.\ D {\bf 78} (2008) 055010
  [arXiv:0712.2039 [hep-ph]].
  %%CITATION = ARXIV:0712.2039;%%
  %155 citations counted in INSPIRE as of 24 Apr 2015

\bibitem{Bertuzzo:2014bwa}
  E.~Bertuzzo, C.~Frugiuele, T.~Gregoire and E.~Ponton,
  %``Dirac gauginos, R symmetry and the 125 GeV Higgs,''
  JHEP {\bf 1504} (2015) 089
  [arXiv:1402.5432 [hep-ph]].
  %%CITATION = ARXIV:1402.5432;%%
  %17 citations counted in INSPIRE as of 24 Apr 2015

%\cite{Diessner:2014ksa}
\bibitem{Diessner:2014ksa}
  P.~Diessner, J.~Kalinowski, W.~Kotlarski and D.~St\"ockinger,
  %``Higgs boson mass and electroweak observables in the MRSSM,''
  JHEP {\bf 1412} (2014) 124
  [arXiv:1410.4791 [hep-ph]].
  %%CITATION = ARXIV:1410.4791;%%
  %6 citations counted in INSPIRE as of 30 Mar 2015
  
%\cite{Diessner:2015yna}
\bibitem{Diessner:2015yna}
  P.~Diessner, J.~Kalinowski, W.~Kotlarski and D.~St\"ockinger,
  %``Two-loop correction to the Higgs boson mass in the MRSSM,''
  arXiv:1504.05386 [hep-ph],
  submitted to Advances in High Energy Physics - Supersymmetry beyond the NMSSM.
  %%CITATION = ARXIV:1504.05386;%%
  
%  \bibitem{LHC_Higgs_ATLAS}
%  The ATLAS collaboration,
%  %``Measurements of the Higgs boson production and decay rates and coupling strengths using pp collision data at √s = 7 and 8 TeV in the ATLAS experiment,''
%  ATLAS-CONF-2015-007, ATLAS-COM-CONF-2015-011.
%  %%CITATION = ATLAS-CONF-2015-007, ATLAS-COM-CONF-2015-011;%%
%  %5 citations counted in INSPIRE as of 24 Apr 2015
%  
%  %\cite{Khachatryan:2014jba}
%\bibitem{Khachatryan:2014jba}
%  V.~Khachatryan {\it et al.}  [CMS Collaboration],
%  %``Precise determination of the mass of the Higgs boson and tests of compatibility of its couplings with the standard model predictions using proton collisions at 7 and 8 TeV,''
%  arXiv:1412.8662 [hep-ex].
%  %%CITATION = ARXIV:1412.8662;%%
%  %62 citations counted in INSPIRE as of 24 Apr 2015

%\cite{Agashe:2014kda}
\bibitem{Agashe:2014kda}
  K.~A.~Olive {\it et al.}  [Particle Data Group Collaboration],
  %``Review of Particle Physics,''
  Chin.\ Phys.\ C {\bf 38} (2014) 090001.
  %%CITATION = CHPHD,C38,090001;%%
  %964 citations counted in INSPIRE as of 24 Apr 2015
  
  \bibitem{degrassi}
G.~Degrassi, S.~Fanchiotti and A.~Sirlin,
  %``Relations Between the On-shell and Ms Frameworks and the M ($W$) - M ($Z$) Interdependence,''
  Nucl.\ Phys.\ B {\bf 351} (1991) 49.
  %%CITATION = NUPHA,B351,49;%%
  %209 citations counted in INSPIRE as of 03 Aug 2014
  
  \bibitem{Peskin:1990zt}
  M.~E.~Peskin and T.~Takeuchi,
  %``A New constraint on a strongly interacting Higgs sector,''
  Phys.\ Rev.\ Lett.\  {\bf 65} (1990) 964.
  %%CITATION = PRLTA,65,964;%%
  %1388 citations counted in INSPIRE as of 11 Sep 2014

\bibitem{Marciano:1990dp}
  W.~J.~Marciano and J.~L.~Rosner,
  %``Atomic parity violation as a probe of new physics,''
  Phys.\ Rev.\ Lett.\  {\bf 65} (1990) 2963
   [Erratum-ibid.\  {\bf 68} (1992) 898].
  %%CITATION = PRLTA,65,2963;%%
  %397 citations counted in INSPIRE as of 07 Oct 2014
  
%\cite{Peskin:1991sw}
\bibitem{Peskin:1991sw}
  M.~E.~Peskin and T.~Takeuchi,
  %``Estimation of oblique electroweak corrections,''
  Phys.\ Rev.\ D {\bf 46} (1992) 381.
  %%CITATION = PHRVA,D46,381;%%
  %1427 citations counted in INSPIRE as of 11 Sep 2014
  
\bibitem{Kennedy:1990ib}
  D.~C.~Kennedy and P.~Langacker,
  %``Precision electroweak experiments and heavy physics: A Global analysis,''
  Phys.\ Rev.\ Lett.\  {\bf 65} (1990) 2967
   [Erratum-ibid.\  {\bf 66} (1991) 395].
  %%CITATION = PRLTA,65,2967;%%
  %297 citations counted in INSPIRE as of 07 Oct 2014
  
\bibitem{Kennedy:1991sn}
  D.~C.~Kennedy and P.~Langacker,
  %``Precision electroweak experiments and heavy physics: An Update,''
  Phys.\ Rev.\ D {\bf 44} (1991) 1591.
  %%CITATION = PHRVA,D44,1591;%%
  %110 citations counted in INSPIRE as of 07 Oct 2014

\bibitem{Altarelli:1990zd}
  G.~Altarelli and R.~Barbieri,
  %``Vacuum polarization effects of new physics on electroweak processes,''
  Phys.\ Lett.\ B {\bf 253} (1991) 161.
  %%CITATION = PHLTA,B253,161;%%
  %631 citations counted in INSPIRE as of 13 Oct 2014
\end{thebibliography}
\end{document}